\documentclass[twocolumn,prl,superscriptaddress]{revtex4-1}
\usepackage{palatino}
\usepackage[latin1]{inputenc}
\usepackage{epsf}
\usepackage{amsmath,amssymb}
\usepackage{latexsym}
\usepackage{calc}
\usepackage{color}
\usepackage{graphicx}
\usepackage{braket}
\usepackage{bm}
\usepackage{xcolor}

\usepackage{layouts}

\begin{document}
\title{Twin-Polaritons: Classical versus Quantum Features in Polaritonic Spectra}
\author{Ir\'en Simk\'o}
\email{is2873@nyu.edu}
\affiliation{Department of Chemistry, New York University, New York, New York 10003, USA}
\affiliation{Simons Center for Computational Physical Chemistry at New York University, New York, New York 10003, USA}
\author{Norah M. Hoffmann}
\email{norah.hoffmann@nyu.edu}
\affiliation{Department of Chemistry, New York University, New York, New York 10003, USA}
\affiliation{Simons Center for Computational Physical Chemistry at New York University, New York, New York 10003, USA}
\affiliation{Department of Physics, New York University, New York, New York 10003, USA}

\date{\today}

\begin{abstract}
Understanding whether a polaritonic phenomenon is fundamentally quantum or classical is essential for building accurate theoretical models and guiding experimental design. Here, we address this question in the context of polaritonic spectra, and report an intriguing new feature: the twin-polariton, an additional splitting beyond the primary resonant polariton splitting, originating from vacuum field fluctuations. We show that the twin-polariton persists in the many-molecule limit under permutationally symmetrical initial-state constraint and that it follows the same linear dependence on coupling strength as the primary polariton splitting. This establishes a novel mechanism by which a quantum feature (the twin-polariton) can be tuned through a classical one (the primary polariton), offering new opportunities to probe and control the fundamental nature of polaritonic systems.
\end{abstract}

\maketitle

Strong light-matter coupling, achievable in optical or plasmonic cavities, forms hybrid light-matter states, called polaritons, and offers a novel way to alter material properties and steer chemical reactions \cite{Ebbesen16,Owrutsky20,Owrutsky21,Ebbesen21B,Owrutsky22}. Due to its potential applications, from manipulating chemical reactivity \cite{Ebbesen16} to room-temperature exciton-polariton condensation that may pave the way for quantum computing devices operating at ambient conditions \cite{peng2022room,georgakilas2025room,ghosh2020quantum}, the number of experimental \cite{Ebbesen16,Ebbesen21A,Ebbesen21B,Xiong24,Borjesson23,Weichman24,Hirai_2020_review,Hirai23, peng2022room,georgakilas2025room,ghosh2020quantum} and theoretical \cite{YuenZhou22,Rubio18,Yelin21,Feist22A, Feist22B, RTFAR18,Flick18,Rubio22,Rubio23,Huo23,Yuen-Zhou18,Nitzan22} studies in polaritonics has rapidly increased in recent years. However, a significant gap remains between experimental reality, where strong coupling requires complex solid-state-like systems (i.e., quantum information devices) or large molecular ensembles (i.e., chemistry), ideally at room temperature, and theoretical models, which struggle with such large-scale calculations. Consequently, conflicting findings persist, highlighting the need for further experimental and theoretical investigation.

In this context, one particularly interesting and often unresolved question is: Which polaritonic phenomena can be fully explained through classical optics, and which require a more advanced quantum  framework?
For instance, the main spectral signature of strong light-matter coupling, the polariton splitting, is often considered a classical feature \cite{WeichmanYuen-Zhou24,welman2025light} and Ref.~\citenum{WeichmanYuen-Zhou24} indicates that many polaritonic phenomena can be explained by polaritons acting as classical optical filters in the $N_{\rm mol} \to \infty$ limit. However, this classical filtering picture may break down \cite{WeichmanYuen-Zhou24} in cases such as polariton-assisted photon recycling \cite{Yuen-Zhou24}, polaritonic 2D IR spectroscopy \cite{Simpkins23} and for systems with many quanta of excitation or few-molecule strong coupling \cite{WeichmanYuen-Zhou24}.

Understanding whether a polaritonic phenomenon is quantum or classical is crucial for developing accurate theoretical models, realizing new experimental designs for quantum computing, or resolving conflicting findings in polaritonic chemistry.
In this paper, we explore this question in the context of polaritonic  spectra, investigating the role of treating the cavity mode classical or quantum mechanically while treating matter quantum mechanically. We report an intriguing new spectral feature whose description requires a quantum treatment of light due to vacuum field fluctuations, and which can persist in the many molecule limit. We call this feature the Twin-Polariton (TP), an additional splitting beyond the primary resonant polariton splitting. Most interestingly, the TP splitting follows the same rule as the primary polariton splitting: a linear increase in splitting with coupling strength, which allows to tune a quantum feature (i.e. the additional TP splitting) using a classical feature (i.e. the primary resonant polariton splitting).

We first define classical and quantum photon modes. A classical photon mode follows Maxwell's equations with boundary conditions (e.g., cavity mirrors), leading to the discretization (or quantization) of light frequencies. However, quantized does not necessarily imply quantum. These modes represent discretized field distributions from boundary conditions but do not account for quantum phenomena such as vacuum field fluctuations or entanglement. This discretization of the electromagnetic field is conceptually similar to Floquet theory, where periodic time dependence leads to discrete quasienergies. However, it is important to emphasize that in our case, the periodicity arises from the spatial boundary conditions of a dark cavity, i.e., a non-driven system, not from a time-periodic external driving field such as a laser. In contrast, in the quantum treatment of light, each quantized photon mode is described by the $\ket{N}$ eigenstates of the photonic Hamiltonian (i.e., Fock states), corresponding to $N$ photons with a specific frequency.

We consider a coupled light-matter system described by the non-relativistic Hamiltonian in the dipole approximation and Coulomb gauge \cite{RFPATR14,FARR17,HARM18,RTFAR18, li2021theory,taylor2020resolution} (using atomic units throughout):
\begin{equation} \label{eq:H_gen}
\hat H = \hat H_{\rm mol} + \hat H_{\rm ph} + g \sqrt{2 \omega_c}\hat{q} \hat \mu + \frac{g^2}{\omega_c} \hat \mu^2. 
\end{equation} 
Here, $\hat H_{\rm mol}$ is the molecular Hamiltonian, and $\hat H_{\rm ph} = (\hat{p}^2 + \omega_c^2 \hat{q}^2)/2$ is the photonic Hamiltonian for a single photon mode with frequency $\omega_c$. The coordinate $\hat{q}$ denotes the photonic displacement-field, and $\hat{p}$ the momentum. The last two terms of Eq.~(\ref{eq:H_gen}) describe the dipole interaction and the dipole self-energy, with dipole moment $\hat \mu$, and coupling strength $g$.

In the \textbf{classical case}, $p(t)$ and $q(t)$ are scalar variables, and the polaritonic spectrum is obtained by time evolving the molecular state $\ket{\Psi(t})$ and the photonic variables \cite{li2022semiclassical}:
\begin{equation} \label{eq:TD_class_1}
    \ket{\dot \Psi(t)} = -i\left(\hat{H}_{\rm mol} + g\sqrt{2\omega_c} q(t) \hat \mu + \frac{g^2}{\omega_c} \hat \mu^2\right)\ket{\Psi(t)}
\end{equation}
\begin{equation} \label{eq:TD_class_2}
    \dot p(t) = -\omega_{\rm c}^2 q(t) - g\sqrt{2\omega_c}\braket{\Psi(t)|\hat \mu|\Psi(t)}
\end{equation}
and $\dot q(t) = p(t)$, where $\ket{\Psi(t)}$ is expressed in the $\ket{\psi_n}$ eigenstates of $\hat{H}_{\rm mol}$. The initial conditions for the photonic vacuum state are $p(0) = q(0) = 0$, and the molecular initial state $\ket{\Psi(0)} = \ket{\psi_i}$. We obtain the absorption spectrum by applying an initial kick and Fourier-transforming the dipole moment \cite{li2022semiclassical} (see SI II.A.). In the \textbf{quantum case}, \(\hat{q}\) and \(\hat{p}\) are expressed in terms of the creation \(\hat{a}^\dagger\) and annihilation \(\hat{a}\) operators, with the light-matter Hamiltonian taking the form:  
\begin{equation} \label{eq:H_QM}  
    \hat{H} = \hat{H}_{\rm mol} + \omega_c \hat{a}^\dagger \hat{a} + g (\hat{a}^\dagger + \hat{a}) \hat{\mu} + \frac{g^2}{\omega_c} \hat{\mu}^2.  
\end{equation}  

The Hamiltonian matrix is constructed using the direct product basis of the molecular eigenstates \(\ket{\psi_n}\) and the photonic number states \(\ket{N}\). In the quantum case, we employ two approaches to obtain the spectrum  (see SI II.B.). The first, commonly used, involves solving the static Schr\"odinger equation by diagonalizing the Hamiltonian to determine the polariton eigenstates. Alternatively, we calculate the spectrum using a time-dependent approach solving the time-dependent Schr\"odinger equation with the initial state \(\ket{\Psi(0)} = \ket{\psi_i,0} = \ket{\psi_i} \ket{0}\). As in the classical model, we apply a kick and obtain the spectrum from the Fourier-transform of the dipole moment. The latter approach is needed for direct comparison with the classical case and to determine time-dependent observables such as populations.

\begin{figure}[h]
    \centering
    \includegraphics[width=\linewidth]{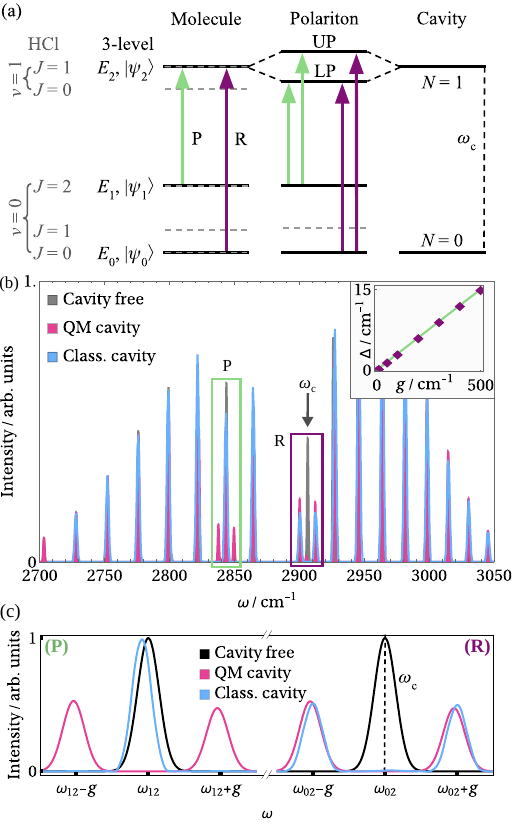}
    \caption{
    (a) Energy-level diagram for a HCl molecule (gray dashed) and a $\Lambda$-type 3-level system (black bold)  coupled to cavity mode $\omega_c$, with the R-branch (purple) and P-branch (green) transitions.  
    (b) Polaritonic rovibrational spectrum of a single HCl, coupled to cavity mode ($\omega_c = 2906.46$ cm$^{-1}$, arrow) with coupling strength $g = 400$ cm$^{-1}$, calculated using the classical (blue) and quantum (pink) cavity models. The cavity-free spectrum is in gray. The inset shows the primary and twin-polariton splittings over coupling strength for the quantum case.  
    (c) Polaritonic spectrum for a single $\Lambda$-type 3-level system, calculated using the classical (blue) and quantum (pink) cavity models.} 
    \label{fig:Fig1}
\end{figure}

To investigate classical versus quantum features in polaritonic spectra, we examine the gas-phase infrared spectra of HCl molecules coupled to a cavity mode, at $T=300$K (see SI I.A. for molecular model) , motivated by recent experiments
reporting strong coupling features in gas-phase systems \cite{Weichman23A,Weichman23B,Weichman24}. This system allows us to analyze individual rotational-vibrational transitions and their modifications under strong coupling. Figure \ref{fig:Fig1}(a) illustrates the rotational-vibrational states of the HCl molecule (left) coupled to a cavity mode polarized in the \(Z\)-direction (right), which is resonant with the \(R\)-branch transition \(\ket{v,J,M} = \ket{0,0,0} \to \ket{1,1,0}\). Here, \(v\) denotes the vibrational quantum number, and \(J\), \(M\) are the rotational quantum numbers. This coupling results in a lower and upper polariton splitting (purple, middle). The corresponding spectrum is shown in Figure \ref{fig:Fig1}(b), where the cavity-free spectrum is plotted in gray, and the spectra for the quantum and classical cavity cases are plotted in pink and blue, respectively. In the \(R\)-branch region (purple box, Fig.~\ref{fig:Fig1}(b)), the polariton splitting is clearly observed in both the classical and quantum calculations, consistent with previous reports that the primary polariton splitting is a classical feature \cite{li2020cavity, WeichmanYuen-Zhou24}. 

The dipole selection rules \(\Delta J = \pm1\) and \(\Delta M = 0\) for light polarized along \(Z\) allow an additional optically active transition \(\ket{v,J,M} = \ket{0,2,0} \to \ket{1,1,0}\) in the \(P\)-branch (green, left and middle of Fig.~\ref{fig:Fig1}(a)). This transition shares the same final state as the \(R\)-branch transition and hence also splits into the upper and lower polariton (green box, Fig.~\ref{fig:Fig1}(b)). 
This TP peak follow the same rule as the primary polariton splitting: the splitting increases linearly with the coupling strength. This is highlighted in the inset of Fig.~\ref{fig:Fig1}(b), where the purple squares represent the Rabi splitting of the primary polariton and the green line represents that of the TP. This tunability allows the off-resonant TP features to be controlled via the primary resonant polariton.

\begin{figure}[h]
    \centering
    \includegraphics[width=\linewidth]{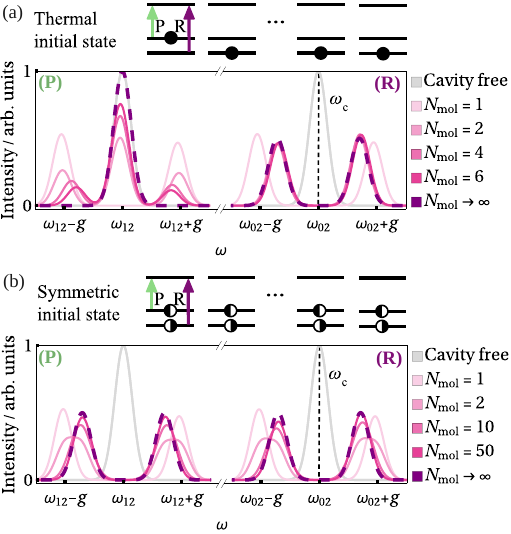}
    \caption{(a) Many-molecule spectrum for a thermal initial state with half the molecules in $\ket{\psi_0}$ and half in $\ket{\psi_1}$, calculated using the quantum cavity model for increasing $N_{\rm mol}$ (pink shades) and the analytic $N_{\rm mol}\!\to\!\infty$ limit (purple, dashed). (b) Same as (a) but for a permutationally symmetric initial state $((\ket{\psi_0}+\ket{\psi_1})/\sqrt{2})^{\otimes N_{\rm mol}}$.}
    \label{fig:many_mol}
\end{figure}

Note that HCl exhibits energy-level characteristics similar to a three-level $\Lambda$-type system \cite{kuhn2002deterministic,mckeever2004deterministic, boozer2007reversible,mucke2010electromagnetically} in atomic cavity QED setups, where  $E_1<<E_2$ and $g<<E_1$ (Fig.~\ref{fig:Fig1}(a), bold), with the corresponding results for the primary polariton and TP are shown in Fig.~\ref{fig:Fig1}(c). The key distinction between molecular polaritonic experiments and atomic cavity QED setups lies in initial state preparation. In HCl, the lower ground-state levels are thermally populated, each weighted by the Boltzmann distribution at 300 K (the standard assumption in spectroscopy), whereas in atomic cavity QED experiments the system can be prepared in various quantum states (e.g., coherent superposition state) via laser excitation. This difference becomes essential in the many-molecule limit (Fig.\ref{fig:many_mol}).

Although the TP phenomenon appears somewhat obvious, as this is true for all optically allowed transitions sharing the same final polariton states, its intriguing aspect lies in its quantum origin. Figure \ref{fig:Fig1}(b,c) (green box) shows that the TP appears only in the quantum calculation, while the spectrum remains unchanged from the cavity-free case in the classical calculation. These findings raise two important questions: (i) Does this feature persist in the many-molecule limit? (ii) Is this feature fundamentally quantum in nature, and if so, what defines its quantum character?

First, we investigate whether the TP persists in the \textbf{many-molecule limit} (see SI II.C). To address this, and to gain physical insight, we employ the simplified $\Lambda$-type three-level model system (Fig.~\ref{fig:Fig1}(a), bold) that reproduces the essential energy level characteristics of HCl. Here we focus on the quantum framework only, necessary to capture the TP, with the normalized coupling $g/\sqrt{N_{\rm mol}}$ and neglecting intermolecular interactions, an approximation appropriate for gas-phase many-molecule experiments or typical $\Lambda$-type setups.

We begin with the thermal ensemble, typical of molecular polaritonics experiments, where the initial state is thermally distributed with $N_0$ molecules in $\ket{\psi_0}$ and $N_1$ molecules in $\ket{\psi_1}$, with $N_{\rm mol}=N_0+N_1$, and no symmetry assumed (Fig.~\ref{fig:many_mol}(a), top). Fig.~\ref{fig:many_mol}(a) shows the TP (left) and primary polariton (right) for the cavity-free case (gray), the quantum case from 1 to 6 molecules (light to dark pink), and the analytic thermodynamic limit $N_{\rm mol}\to\infty$ (purple) (see SI III.). As the number of molecules increases, the TP feature weakens and a peak emerges at the cavity-free frequency. In the thermodynamic limit, the TP vanishes entirely. This peak arises from transitions into dark states, which are forbidden for the R transition but allowed for the P transition. Since the total dipole operator is symmetric under permutation of molecules, a thermal permutation-nonsymmetric initial state permits transitions to both bright symmetric polaritonic states (split peaks of primary polariton and TP) and dark nonsymmetric states (central peak of TP), leading to the TP vanishing in the many-molecule limit.

Hence for the TP to persist, the initial state must be a permutationally symmetric state, e.g., $((\ket{\psi_0}+\ket{\psi_1})/\sqrt{2})^{\otimes N_{\rm mol}}$ (Fig.~\ref{fig:many_mol}(b), top). Such symmetry ensures that only symmetric final states are accessible, restricting transitions to the bright polaritonic manifold and suppressing dark-state absorption. Such initial states can be prepared by lasers in cold-atom experiments, in Bose-Einstein condensates, or enforced by the Pauli principle for bosonic particles in a cavity when particles can be interchanged \cite{Szidarovszky23A,Szidarovszky21}.
Figure~\ref{fig:many_mol}(b) shows the spectrum of the model system with such a symmetric initial state for the TP (left) and the primary polariton (right), comparing the cavity-free case (gray), increasing numbers of molecules from 1 to 50 (light to dark pink), and the analytical thermodynamic limit (purple) (SI~III.). Here we find that indeed, as the number of molecules increases, the TP persists (left). Note, this symmetry constraint does not mean that the TP cannot survive under ambient conditions. If a symmetric initial state can be maintained at room temperature (e.g., in polariton Bose-Einstein condensates), the TP will remain even in the many-molecule/atom limit. Hence that the TP survives only under these symmetry constraints underscores its quantum character, since permutationally symmetric states represent coherent superpositions across atoms, a main characteristic of quantum light-matter interactions.

\begin{figure}
    \centering
    \includegraphics[width=\linewidth]{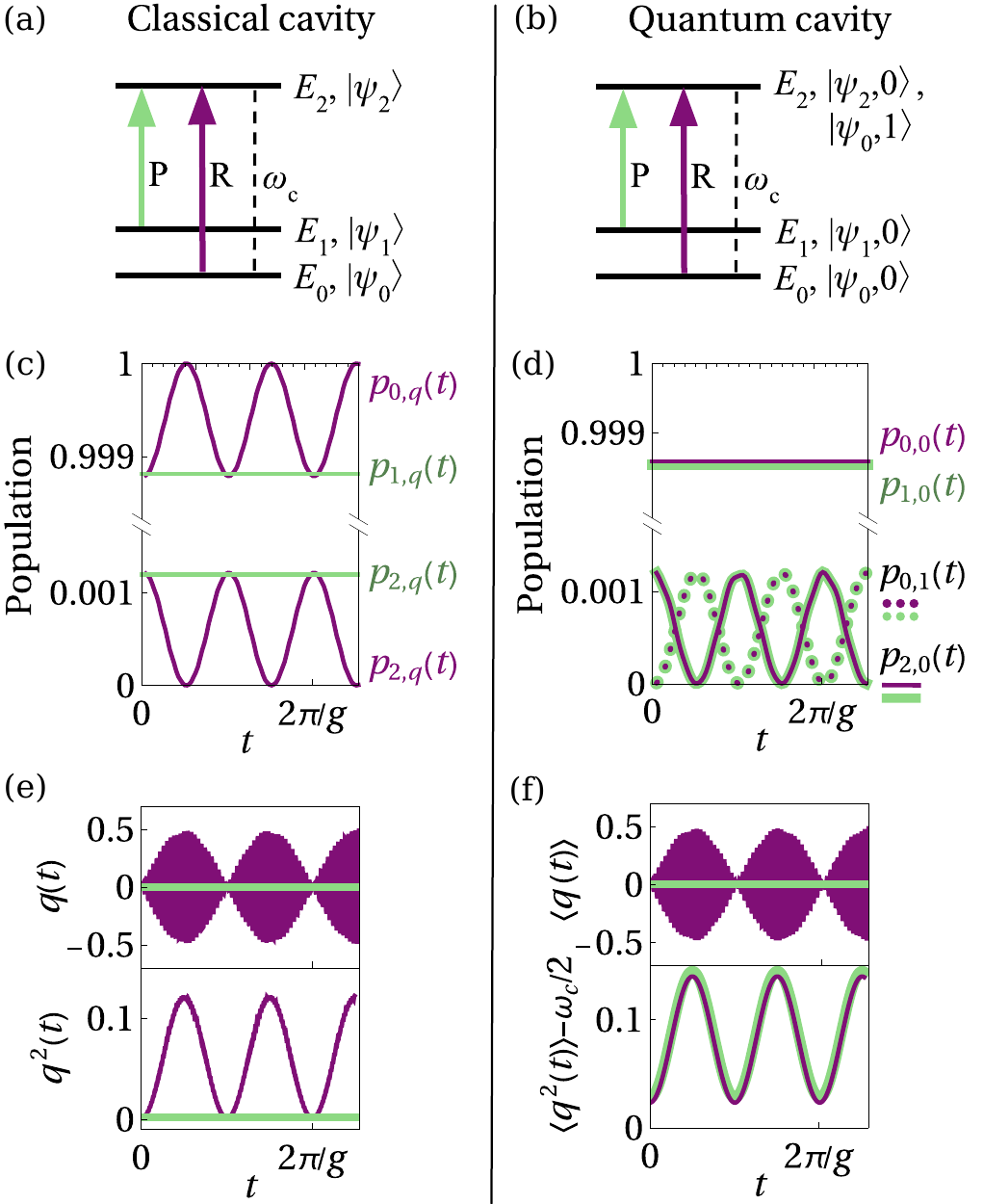}
    \caption{
    (a, b) Diagram of the basis states and transitions in the classical (a) and quantum (b) cavity model.
    (c, d) Population analysis for the time-dependent classical (c) and quantum (d) cavity simulations, where the initial molecular state was either a perturbed $\ket{\psi_0}$ (R-branch, purple) or $\ket{\psi_1}$ (P-branch, green). 
    Electric field displacement $q(t)/\braket{q(t)}$ and its intensity $q^2(t)/\braket{q^2(t)}$ from classical(e)/quantum(f) cavity model.}
    \label{fig:QM_vs_class}
\end{figure}

Having confirmed that the TP feature persists in the many-molecule case under symmetry constraints, we now turn to the question of whether it constitutes an actual \textbf{quantum effect} and, if so, what defines its quantum character. To address this, we analyze the time-dependent population dynamics of our three-level system in both classical and quantum descriptions, for a single molecule. In the classical case, the cavity-coupled molecular wavefunction \( \ket{\Psi(t)}\) is given by  
\begin{equation}  
\ket{\Psi(t)} = \sum_{k} C_{k,q}(t) e^{-iE_{k} t} \ket{\psi_{k}},  
\end{equation}
and the corresponding population of the $\ket{\psi_{k}}$ state is $p_{k,q}(t) = |C_{k,q}(t)|^2$,
where \( k \) denotes the molecular state and \( q \) accounts for the classical photon mode dependence of the wavefunction, obtained by solving Eqs.(\ref{eq:TD_class_1}-\ref{eq:TD_class_2}). In the quantum case, the wavefunction and the corresponding populations are defined as  
\begin{equation}  
\ket{\Psi(t)} = \sum_{N} \sum_{k} C_{k,N}(t) e^{-i(E_{k} +N\omega_c)t} \ket{\psi_{k},N},  
\end{equation} 
and $p_{k,N}(t) = |C_{k,N}(t)|^2$, where \( N \) denotes the photon Fock state. The time-dependent populations for both the classical and quantum cases are plotted in Fig.~\ref{fig:QM_vs_class}(c,d) respectively. Furthermore, the dipole moment can be expressed as  
\begin{equation}
    \braket{\mu(t)}\!\!=\!\!\sum_N \sum_{k,l} C^*_{l,N}(t)C_{k,N}(t)e^{-i(E_l-E_k)t}\!\braket{\psi_{l}|\hat{\mu}|\psi_k}.
\end{equation}
in the quantum case and as 
\begin{equation}
    \braket{\mu(t)}= \sum_{k,l} C^*_{l,q}(t)C_{k,q}(t)e^{-i(E_l-E_k)t}\braket{\psi_{l}|\hat{\mu}|\psi_k}
\end{equation}
in the classical case, where $k,l$ denote molecular states. This establishes a direct connection between population transfer and spectral features via the Fourier transform of the time-dependent dipole moment. In other words, an oscillatory population transfer with frequency \( \Omega \) leads to a peak splitting with a separation of \( \hbar\Omega \), whereas a constant population results in a single peak.

Fig.~\ref{fig:QM_vs_class}(c) presents the classical population dynamics of the three-level system, 
with the R-branch transition shown in purple and the P-branch in green. In the classical case, we observe that only the R-branch transition exhibits oscillatory populations $p_{0,q}(t)$ and $p_{2,q}(t)$, indicating oscillatory energy transfer between the molecule and the cavity mode. The population oscillation results in the expected primary resonant polariton peak splitting due to cavity mode coupling. In contrast, for the P-branch, the populations $p_{1,q}(t)$ and $p_{2,q}(t)$ remain constant, explaining the absence of TP splitting in the classical calculation. Turning to the quantum case, depicted in Fig.~\ref{fig:QM_vs_class}(d), we find a striking difference: \textbf{both} the R-branch and P-branch transitions exhibit oscillatory populations $p_{2,0}(t)$  and $p_{0,1}(t)$ (corresponding to the \( \ket{\psi_2,0} \) and \( \ket{\psi_0,1} \) basis states). This oscillation ultimately gives rise to both the main and TP splittings.

While the quantum case reveals that population oscillations are essential for the TP, the next question is what initiates this transfer and why it is missing from the classical description. In the classical model, population transfer requires nonzero $q(t)$ (see Eqs.~(\ref{eq:TD_class_1})-(\ref{eq:TD_class_2})), but since the initial state is $p(0)=q(0)=0$, nonzero $q(t)$ can be only induced if the dipole moment $\braket{\Psi(t)|\hat \mu|\Psi(t)}$ oscillates with $\omega_c$, which is satisfied for the R transition but not for P. This suggest that the distinguishing feature absent in the classical photon-mode description is the vacuum field fluctuation. To investigate this, we analyze the time evolution of the photonic displacement field $q(t)$ and its intensity $q^2(t)$ for the R- and P-branch transitions in both classical and quantum models. For the latter we calculate $\braket{q(t)}$ and $\braket{q^2(t)}$, but omit  the bracket for clarity. In the classical case, the R transition (Fig.~\ref{fig:QM_vs_class}(e), purple) couples to the cavity, yielding nonzero oscillating $q(t)$ and driving the population transfer responsible for the primary polariton splitting. In contrast, the P transition (green) does not couple effectively, resulting in $q(t)=0$. The same behavior is seen in the quantum case (Fig.~\ref{fig:QM_vs_class}(f)). However, a crucial difference emerges when examining the intensity $q^2(t)$. In the classical model, $q^2(t)$ is simply the square of $q(t)$, so it vanishes for the P transition and cannot induce splitting. In the quantum model, by contrast, $q^2(t)$ is oscillatory for the P transition even though $q(t)=0$ is constant, directly reflecting vacuum field fluctuations. These fluctuations initiate the population transfer that gives rise to the TP. Thus, the TP is fundamentally a vacuum-induced effect and can be viewed as a vacuum-induced Autler-Townes splitting \cite{ding2018vacuumATS}, where the key distinction from the traditional Autler-Townes effect \cite{AutlerTownes55,cohen96} is that here the R transition (the primary polariton) is not driven by an external strong laser but by the cavity mode itself, which governs both the primary polariton and the TP.

In conclusion, we investigated classical versus quantum features in the polaritonic rovibrational spectrum of gas-phase molecules and $\Lambda$-type systems. We identified an intriguing new feature, the Twin-Polariton (TP), arising from vacuum field fluctuations. The TP persists in the many-molecule limit under initial state symmetry constraints, such as those possible in Bose-Einstein condensates and cold atom setups. Perhaps most remarkable is that the TP appears alongside the primary polariton splitting and follows the same linear dependence on coupling strength, enabling a quantum feature (the TP) to be tuned via a classical one (the primary splitting) using only the cavity mode, with no additional laser source required. Moreover, the effect is reversible: when the cavity is resonant with the P-branch, the TP emerges in the R-branch, provided both transitions are optically allowed and share the same final state. This provides a novel lever for manipulating molecular polaritonic systems and offers new insights into their fundamental nature. Because of its off-resonant character, we hypothesize that experimental observation of the TP will require absorption measurements along the non-confined cavity direction, and may be feasible in cold-atom setups or even room-temperature (polariton) Bose-Einstein condensates that satisfy the necessary symmetry constraints.

\section*{Acknowledgements}This work was supported by the Simons Center for Computational Physical Chemistry (SCCPC) at NYU (SF Grant No. 839534). Additional support was provided in part through NYU IT High Performance Computing resources, services, and staff expertise. Simulations were partially executed on resources supported by the SCCPC. I.S. acknowledges support from a postdoctoral fellowship awarded by the SCCPC at NYU.

\bibliography{bibliogra}

\end{document}


\title{Supplementary Information}
\author{Ir\'en Simk\'o}
\email{is2873@nyu.edu}
\affiliation{Department of Chemistry, New York University, New York, New York 10003, USA}
\affiliation{Simons Center for Computational Physical Chemistry at New York University, New York, New York 10003, USA}
\author{Norah M. Hoffmann}
\email{norah.hoffmann@nyu.edu}
\affiliation{Department of Chemistry, New York University, New York, New York 10003, USA}
\affiliation{Simons Center for Computational Physical Chemistry at New York University, New York, New York 10003, USA}
\affiliation{Department of Physics, New York University, New York, New York 10003, USA}

\maketitle


\section{Molecular Models}
\subsection{HCl}
 For the HCl molecule calculations we employ the following spectroscopically accurate Morse potential energy curve:
\begin{equation}
    V(R) = D_e(e^{-2\alpha(R-R_e)}-2e^{-\alpha(R-R_e)}+1),
\end{equation}
where $R_e=2.40855$ bohr is the equilibrium bond length, $D_e=37209.369$ cm$^{-1}$ is the binding energy, and $\alpha=0.993099$ bohr. These parameters were obtained by refitting the spectroscopically accurate potential in Ref.\cite{coxon2000radial}, which itself is fitted to measured spectra. The dipole moment curve is taken from Ref.\cite{Maroulis91}, and the masses are m(H)=1837.1522 a.u. and  m(Cl)=63744.3019 a.u.. In the case of the HCl calculation, we assume that both the cavity light mode and the laser used to measure the spectrum are polarized in the $Z$ direction. Therefore, there is no interaction or transition between states with different $M$ rotational quantum number. 

The cavity is  coupled to the \(\ket{v,J,M} = \ket{0,0,0} \to \ket{1,1,0}\) transition with $\omega_c = 2906.46$ cm$^{-1}$ frequency, while the coupling strength was set to $g=400$ cm$^{-1}$. The calculated spectrum corresponds to 300K, obtained from one-molecule spectra with the different initial states weighted by the Boltzmann-distribution.
We emphasize that the purpose of this calculation is illustrative, hence the temperature and cavity parameters differ from those in current experimental setups \cite{Weichman23A,Weichman23B,Weichman24}, and collective coupling is not included in this calculation.

\subsection{3-level system}
To keep the discussion general, results in the main text are presented in relation to parameters (i.e., cavity frequency and coupling strength) instead of specific values. In practice, computations for the 3-level model system were performed using the following parameter values: energies are given by $E_0 = 0$ a.u., $E_1 = 2\cdot 10^{-3}$ a.u., $E_2 = 10\cdot 10^{-3}$ a.u., the coupling strength was set to $g=0.2\cdot 10^{-3}$ a.u., and dipole matrix elements are $\mu_{02} = \mu_{20} = \mu_{12} = \mu_{21} =1$ a.u. and $\mu_{ij}=0$ otherwise. The results for the 3-levels system are presented parametrically instead of specific numbers. The results correspond to parameter sets where $E_1<<E_2$ and $g<<E_1$.

\section{Additional computational details}

In the following, we provide a detailed discussion of the calculations for both the classical and quantum light models. It is important to note that in both cases, the material system is treated fully quantum mechanically, while only the treatment of light differs. The corresponding Hamiltonians can be found in Eqs. (1) and (4) of the main text. We only provide general expressions,  assuming the cavity-free eigenstates have been already computed. 
Detailed expressions for matrix elements involving rovibrational states can be found, for example, in Ref.\cite{Szidarovszky23}.

Additionally, in the fully quantum case, we consider two different approaches for computing observables: a static and a dynamic framework. This distinction allows for a direct comparison with classical observables. A key difference between the static quantum and time-dependent quantum (or classical) cavity models lies in the interpretation of polaritons. In the static framework, polaritons correspond to eigenstates of $\hat{H}$ with hybrid light-matter character, where the degeneracy of a molecular excited state and a photonic excited state is lifted due to light-matter coupling. In contrast, in the time-dependent framework, polaritons emerge as dynamical features arising from periodic energy exchange between the molecule and the cavity light  mode.

\subsection{Quantum molecule - classical light mode model}
In the classical light model, $p$ and $q$ are scalar variables. The initial conditions for the photonic vacuum state are $p(0) = q(0) = 0$ and the molecular initial state is an eigenstate of $\hat{H}_{\rm mol}$, given by $\ket{\Psi(0)} = \ket{\psi_i}$. The time evolution of the light-matter system is governed by the following equations of motion:

\begin{align} \label{eq:TD_class}
    \ket{\dot \Psi(t)} &= -i \left(\hat{H}_{\rm mol} + (g\sqrt{2\omega_c} q(t) + f(t))\hat \mu + \frac{g^2}{\omega_c} \hat \mu^2\right)\ket{\Psi(t)}, \\
    \dot p(t) & = -\omega_{\rm c}^2 q(t) - g\sqrt{2\omega_c}\braket{\Psi(t)|\hat \mu|\Psi(t)}, \\
    \dot q(t) &= p(t), 
\end{align}
where we apply a sharp but weak Gaussian pulse $f(t)$ as an initial kick to move the system out of equilibrium. The time-dependent molecular state is expressed in the basis of the cavity-free molecular eigenstates $ \ket{\psi_k} $ with

\begin{equation} \label{eq:psi_t_class}
    \ket{\Psi(t)} = \sum_k C_{k,q}(t)e^{-iE_kt}\ket{\psi_k},
\end{equation}

where the subscript $q$ indicates that the coefficients depend on the photonic state. The population of $\ket{\psi_k}$ is given by $p_{k,q}(t) = |C_{k,q}(t)|^2$, where the subscript $q$ indicates that the molecular state depends on the photonic coordinate. Substituting Eq. (\ref{eq:psi_t_class}) in Eq. (\ref{eq:TD_class}), we obtain the equations of motion expressed in terms of the $C_{k,q}(t)$ coefficients:

\begin{align} \label{eq:TD_class_mtx}
    \dot C_{n,q}(t) &= iE_n C_{n,q}(t) - i \sum_k C_{k,q}(t)e^{-i(E_k-E_n)t} \Big( E_n\delta_{n,k} + (g\sqrt{2\omega_c} q(t) + f(t))\braket{\psi_n|\hat \mu|\psi_k} 
    + \frac{g^2}{\omega_c} \braket{\psi_n|\hat \mu^2|\psi_k} \Big), \nonumber \\
    \dot p(t) &= -\omega_{\rm c}^2 q(t) - g\sqrt{2\omega_c} \braket{\mu(t)}, \\
    \dot q(t) &= p(t),
\end{align}

where 

\begin{equation}
    \braket{\mu(t)} = \braket{\Psi(t)|\hat \mu|\Psi(t)} = \sum_{k,k'} C^*_{k',q}(t) C_{k,q}(t)e^{-i(E_k-E_{k'})t} \braket{\psi_{k'}|\hat \mu|\psi_k},
\end{equation}

is the time-dependent expectation value of the dipole moment. The matrix element of the $\hat \mu^2$ is calculated by inserting a resolution of identity:
\begin{equation}\label{eq:mu_squared}
    \braket{\psi_n|\hat \mu^2|\psi_{n'}} =\sum_k \braket{\psi_n|\hat \mu|\psi_k}\braket{\psi_k|\hat \mu|\psi_{n'}},
\end{equation}
where the sum goes over all the basis states, i.e., the 3 states for the model system and the $J=0-10$ states in the $v=0$ and 1 vibrational manifolds for the HCl. Essentially the matrix of $\hat \mu^2$ is approximated as the square of the $\hat \mu$ matrix. We note that if the calculation is sensitive to the dipole self-energy, excited electronic states may need to be included in the sum in Eq.~(\ref{eq:mu_squared}) \cite{borges2024extending,fabri2025impact}. However, for our results, the dipole self-energy has no significant effect and can therefore be neglected.

We obtain the absorption spectrum associated with the initial state $\ket{\psi_i}$ from the Fourier-transform of $\braket{\mu(t)}$\cite{li2022semiclassical} as
\begin{equation}\label{eq:TD_int_i}
    I_i(\omega) \propto \omega^2 \int_{-\infty}^{\infty} e^{-i \omega t}\braket{\mu(t)}dt.
\end{equation}

\subsection{Full quantum model}
In the quantum case, $\hat{q}$ and $\hat{p}$ are expressed in terms of the creation $\hat{a}^\dagger$ and annihilation $\hat{a}$ operators,
\begin{equation}
    \hat q = \sqrt{\frac{1}{2\omega_c}}(\hat{a}^\dagger + \hat{a})
    \qquad \text{and} \qquad
    \hat p = i\sqrt{\frac{\omega_c}{2}}(\hat{a}^\dagger - \hat{a}).
\end{equation}

Then, the dipole interaction term and the dipole self-energy are 
\begin{equation} \label{eq:Vdip_Vdse}
    \hat{V}_{\rm dip} = g (\hat{a}^\dagger + \hat{a}) \hat{\mu}
    \qquad \text{and} \qquad
    \hat{V}_{\rm dse} = \frac{g^2}{\omega_c} \hat{\mu}^2,
\end{equation}
respectively.

To obtain the polaritonic energies and eigenstates, we solve the time-independent Schr{\"o}dinger-equation by diagonalizing the matrix of $\hat H$ in the $\ket{\psi_k,N}=\ket{\psi_k}\ket{N}$ basis, where $\ket{N}$ denotes photon number states (Fock states). The matrix elements are given by
\begin{align}
    \braket{\psi_{k'}, N'|\hat{H}_{\rm mol}| \psi_{k}, N} &= \delta_{N',N} \delta_{k',k}E_k, \\
    \braket{\psi_{k'}, N'|\hat{H}_{\rm ph}| \psi_{k}, N} &= \delta_{N',N} \delta_{k',k} N \omega_c, \\
    \braket{\psi_{k'}, N'|\hat{V}_{\rm dip}| \psi_{k}, N} &= g \braket{\psi_{k'}, N'|(\hat{a}^\dagger + \hat{a}) \hat{\mu}| \psi_{k}, N} = g (\sqrt{N+1}\delta_{N', N+1}+\sqrt{N}\delta_{N', N-1})\braket{\psi_{k'}|\hat{\mu}| \psi_{k}}, \\
    \braket{\psi_{k'}, N'|\hat{V}_{\rm dse}| \psi_{k}, N} &= \frac{g^2}{\omega_c} \braket{\psi_{k'}, N'| \hat{\mu}^2| \psi_{k}, N} = \frac{g^2}{\omega_c} \delta_{N',N}\braket{\psi_{k'}| \hat{\mu}^2| \psi_{k}}.
\end{align}

\subsubsection{Spectrum from the static framework}
The eigenstates of $\hat H$ expressed in the $\ket{\psi_k,N}$ basis are
\begin{equation}
    \ket{\Psi} = \sum_{k} \sum_{N} C_{k,N}\ket{\psi_k,N}.
\end{equation}
The absorption spectrum can be directly calculated from the polariton eigenstates, with the intensity of the transition between $\ket{\Psi_i}$ and $\ket{\Psi_f}$ is  $I(\Psi_i \to \Psi_f) \propto|\braket{\Psi_i|\hat \mu|\Psi_f}|^2$.

\subsubsection{Spectrum from the time-dependent framework}
Alternatively, the spectrum can be obtained using the time-dependent framework in the quantum model. This approach enables direct comparison with the classical light model. Here, we solve the time-dependent Schr{\"o}dinger equation
\begin{equation}
    \ket{\dot \Psi(t)} = -i(\hat H + f(t) \hat{\mu})\ket{\Psi(t)}
\end{equation}
where, as in the classical light model, a sharp but weak Gaussian pulse $f(t)$ is initially applied, and the initial condition is $\ket{\Psi(0)}=\ket{\psi_i,0}$. The time-dependent wave function of the coupled light-matter system is expressed as 

\begin{equation}
    \ket{\Psi(t)} = \sum_{k} \sum_{N} C_{k,N}(t)e^{-i(E_k+N\omega_c)t}\ket{\psi_k,N},
\end{equation}
while the dipole moment is 
\begin{equation}
    \braket{\mu(t)}=\sum_N \sum_{k,k'} C^*_{k',N}(t)C_{k,N}(t)e^{-i(E_k-E_{k'})t}\braket{\psi_{k'}|\hat{\mu}|\psi_k}.
\end{equation}
The spectrum is then computed from $\braket{\mu(t)}$ according to Eq.~(\ref{eq:TD_int_i}).

\subsection{Many-molecule case}
The many-molecule case is treated the same way as the single-molecule case in both the quantum and classical models, except that the operators and wave functions of the material now correspond to many molecules and the coupling strength is scaled as $g/\sqrt{N_{\rm mol}}$. The inter-molecular interactions are neglected, therefore the many-molecule  molecular Hamiltonian is the sum of single-molecule Hamiltonians: ${\hat H}_{\rm mol} = \sum_{k=1}^{N_{\rm mol}}{\hat H}_{\rm mol}(k)$, while the many-molecule dipole moment is the sum of the dipoles of the individual molecules: $\bm{\hat \mu} = \sum_{k=1}^{N_{\rm mol}}\bm{\hat \mu}(k)$. The many-molecule basis states are the product of single-molecule states: $\ket{{i_1}...i_{N_{\rm mol}}} = \ket{{i_1}(1)}...\ket{i_{N_{\rm mol}}(N_{\rm mol})}$,
where $\ket{{i_k}(k)}$ means that the $k$th molecule is in the $\ket{\psi_{i_k}}$ state.

\section{Analytical derivations for the many-molecule case}
\subsection{Eigenstates of the coupled light-matter Hamiltonian }
In this section we derive approximate analytical formulas for the polariton states and polariton spectrum in the  $N_{\rm mol}\to \infty$ limit. For now, we neglect the dipole self-energy in the Hamiltonian and scale the coupling strength as $g/\sqrt{N_{\rm mol}}$, and as always, we assume that the cavity frequency is $\omega_c=E_2-E_0$:
\begin{equation} \label{eq:H_QM}  
    \hat{H} = \hat{H}_{\rm mol} + \omega_c \hat{a}^\dagger \hat{a} + \hat{V}_{\rm dip}.  
\end{equation}  
Here, $\hat{V}_{\rm dip}=g/\sqrt{N_{\rm mol}} (\hat{a}^\dagger + \hat{a}) \hat{\mu}$ is the dipole coupling term, ${\hat H}_{\rm mol} = \sum_{k=1}^{N_{\rm mol}}{\hat H}_{\rm mol}(k)$ is the the many-molecule molecular Hamiltonian, and the many-molecule dipole moment is approximated as the sum of the dipoles of the individual molecules, $\bm{\hat \mu} = \sum_{k=1}^{N_{\rm mol}}\bm{\hat \mu}(k)$. In order to construct the Hamiltonian matrix, we denote the  $N_{\rm mol}$-molecule states as
\begin{equation}
    \ket{{i_1}...i_{N_{\rm mol}}} = \ket{{i_1}(1)}...\ket{i_{N_{\rm mol}}(N_{\rm mol})},
\end{equation}
where $\ket{{i_k}(k)}$ means that the $k$th molecule is in the $\ket{\psi_{i_k}}$ state.
The $\ket{{i_1}...i_{N_{\rm mol}}}\ket{N}$ basis states, where all molecules are in $\ket{\psi_0}$ or $\ket{\psi_1}$ and the cavity is in the vacuum state belong to the ground state manifold, while in the first excited manifold, either the cavity is in $\ket{N=1}$ or exactly one molecule is in $\ket{\psi_2}$. Assuming the ground state manifold is unaffected by light-matter coupling, we construct the Hamiltonian matrix in the first excited manifold. Let us define a molecular state in the ground state manifold as
\begin{equation}
    \ket{G^{(1)}(n_0,n_1=N_{\rm mol}-n_0,n_2=0)}\equiv \ket{\overbrace{0...0}^{n_0}\overbrace{1...1}^{N_{\rm mol}-n_0}},
\end{equation}
where $n_i$ in the parentheses denotes the occupation of $\ket{\psi_i}$. Note that there are ${N_{\rm mol} \choose n_0}$ states with the same occupation numbers, denoted by $\ket{G^{(k)}(n_0,N_{\rm mol}-n_0,0)}$, where $k=1,...,{N_{\rm mol} \choose n_0}$. We also define excited states originating from $\ket{G^{(1)}(n_0,N_{\rm mol}-n_0,0)}$, where exactly one molecule has been excited from $\ket{\psi_0}$ to $\ket{\psi_2}$. The symmetric excited state is defined as 
\begin{equation}
    \ket{S^{(1)}(n_0-1,N_{\rm mol}-n_0,1)}\equiv 
    \frac{1}{\sqrt{n_0}}\sum_{i=1}^{n_0}\ket{\overbrace{0...\underset{i}{2}...0}^{n_0}\overbrace{1...1}^{N_{\rm mol}-n_0}},
\end{equation}
while the $n_0-1$ dark excited states are
\begin{equation}
    \ket{D^{(1)}_k(n_0-1,N_{\rm mol}-n_0,1)}\equiv \sum_{i=1}^{n_0}c^{(1)}_{k,i}\ket{\overbrace{0...\underset{i}{2}...0}^{n_0}\overbrace{1...1}^{N_{\rm mol}-n_0}},
\end{equation}
where $k=1,..., n_0-1$. The dark states are orthonormal, and also orthogonal to $\ket{S^{(j)}(n_0-1,N_{\rm mol}-n_0,1)}$, therefore $\sum_{i=1}^{n_0}c^{(j)}_{k,i}=0$. In total, there are ${N_{\rm mol} \choose n_0-1}$ molecular states with $(n_0,n_1,n_2)=(n_0-1,N_{\rm mol}-n_0,1)$ occupation numbers, obtained from all possible $\ket{G^{(j)}(n_0,N_{\rm mol}-n_0,0)}$ states.

In order to calculate the transition dipole matrix elements, it is useful to note that 
\begin{equation}
    \braket{i'_1...i'_{N_{\rm mol}}|\hat \mu|{i_1}...i_{N_{\rm mol}}} = \sum_{k=1}^{N_{\rm mol}} \braket{i'_k(k)|\hat \mu|i_k(k)}\delta_{i'_1,i_1}...\delta_{i'_{k-1},i_{k-1}}\delta_{i'_{k+1},i_{k+1}}\delta_{i'_{N_{\rm mol}},i_{N_{\rm mol}}},
\end{equation}
which means that $\braket{{i'_1}...i'_{N_{\rm mol}}|\hat \mu |{i_1}...i_{N_{\rm mol}}}$ is nonzero only if the $({i'_1}...i'_{N_{\rm mol}})$ and $({i_1}...i_{N_{\rm mol}})$ quantum number strings differ for \emph{exactly one} molecule. Therefore, 
$\ket{D^{(1)}_k(n_0-1,N_{\rm mol}-n_0,1)}$ and $\ket{S^{(1)}(n_0-1,N_{\rm mol}-n_0,1)}$ may have nonzero dipole matrix elements only with states with $(n_0,n_1,n_2)=(n_0,N_{\rm mol}-n_0,0)$ occupation numbers:
\begin{equation}
    \braket{G^{(1)}(n_0,N_{\rm mol}-n_0,0)|\hat{\mu}|S^{(1)}(n_0-1,N_{\rm mol}-n_0,1)}=
    \frac{1}{\sqrt{n_0}}\sum_{i=1}^{n_0}\braket{\overbrace{0...0}^{n_0}\overbrace{1...1}^{N_{\rm mol}-n_0}|\hat{\mu}|\overbrace{0...\underset{i}{2}...0}^{n_0}\overbrace{1...1}^{N_{\rm mol}-n_0}}=\sqrt{n_0}\mu,
\end{equation}
\begin{equation}
    \braket{G^{(1)}(n_0,N_{\rm mol}-n_0,0)|\hat{\mu}|D^{(1)}_k(n_0-1,N_{\rm mol}-n_0,1)}=
    \sum_{i=1}^{n_0}c^{(1)}_{k,i}\braket{\overbrace{0...0}^{n_0}\overbrace{1...1}^{N_{\rm mol}-n_0}|\hat{\mu}|\overbrace{0...\underset{i}{2}...0}^{n_0}\overbrace{1...1}^{N_{\rm mol}-n_0}}=0,
\end{equation}
or $(n_0-1,N_{\rm mol}-n_0+1,0)$ occupation numbers:
\begin{equation}
    \braket{\overbrace{0...\underset{j}{1}...0}^{n_0}\overbrace{1...1}^{N_{\rm mol}-n_0}|\hat{\mu}|S^{(1)}(n_0-1,N_{\rm mol}-n_0,1)}=
    \frac{1}{\sqrt{n_0}}\sum_{i=1}^{n_0}\braket{\overbrace{0...\underset{j}{1}...0}^{n_0}\overbrace{1...1}^{N_{\rm mol}-n_0}|\hat{\mu}|\overbrace{0...\underset{i}{2}...0}^{n_0}\overbrace{1...1}^{N_{\rm mol}-n_0}}=\frac{\mu}{\sqrt{n_0}},
\end{equation}
\begin{equation}
    \braket{\overbrace{0...\underset{j}{1}...0}^{n_0}\overbrace{1...1}^{N_{\rm mol}-n_0}|\hat{\mu}|D^{(1)}_k(n_0-1,N_{\rm mol}-n_0,1)}=
    \sum_{i=1}^{n_0}c^{(1)}_{k,i}\braket{\overbrace{0...\underset{j}{1}...0}^{n_0}\overbrace{1...1}^{N_{\rm mol}-n_0}|\hat{\mu}|\overbrace{0...\underset{i}{2}...0}^{n_0}\overbrace{1...1}^{N_{\rm mol}-n_0}}=c^{(1)}_{k,j} \mu .
\end{equation}

From this, the Hamiltonian matrix can now be constructed in the first excited manifold. The matrices of the uncoupled molecular and photonic Hamiltonians, $\hat{H}_{\rm mol}+\omega_c \hat{a}^\dagger \hat{a}$, are diagonal. The diagonal matrix element associated with the  $\ket{S^{(l)}(n_0-1,N_{\rm mol}-n_0,1}\ket{0}$, $\ket{D^{(m)}_k(n_0-1,N_{\rm mol}-n_0,1}\ket{0}$, and $\ket{G^{(n)}(n_0,N_{\rm mol}-n_0,0)}\ket{1}$ basis functions is $E(n_0-1,N_{\rm mol}-n_0,1)$, where $E(n_0,n_1,n_2)\equiv n_0E_0+n_1E_1+n_2 E_2$.

Using the dipole matrix elements calculated above, we obtain matrix elements of $\hat{V}_{\rm dip}$. The following offdiagonal matrix elements connect degenerate basis states with $E(n_0-1,N_{\rm mol}-n_0,1)$ energy:
\begin{align}\label{eq:mu_mtx_1}
    \bra{1}\braket{G^{(1)}(n_0,N_{\rm mol}-n_0,0)|\hat{V}_{\rm dip}|S^{(1)}(n_0-1,N_{\rm mol}-n_0,1)}\ket{0}&=g\sqrt{\frac{n_0}{N_{\rm mol}}}\mu\\
    \bra{1}\braket{G^{(1)}(n_0,N_{\rm mol}-n_0,0)|\hat{V}_{\rm dip}|D^{(1)}_k(n_0-1,N_{\rm mol}-n_0,1)}\ket{0}&=0.
\end{align}
While the following matrix elements connect nondegenerate basis states, with $E(n_0-1,N_{\rm mol}-n_0,1)$ and $E(n_0-2,N_{\rm mol}-n_0+1,1)$ energies:
\begin{align} \label{eq:mu_mtx_2}
    \bra{1}\braket{\overbrace{0...\underset{j}{1}...0}^{n_0}\overbrace{1...1}^{N_{\rm mol}-n_0}|\hat{V}_{\rm dip}|S^{(1)}(n_0-1,N_{\rm mol}-n_0,1)}\ket{0}&=
    \frac{g}{\sqrt{n_0 N_{\rm mol}}}\mu \\
    \bra{1}\braket{\overbrace{0...\underset{j}{1}...0}^{n_0}\overbrace{1...1}^{N_{\rm mol}-n_0}|\hat{V}_{\rm dip}|D^{(1)}_k(n_0-1,N_{\rm mol}-n_0,1)}\ket{0}&=c^{(1)}_{k,j}\frac{g}{\sqrt{N_{\rm mol}}} \mu .
\end{align}
In this analytical derivation, we neglect the offdiagonal matrix elements that connect nondegenarate basis states. (Note that all matrix elements are kept in the numerical code.) Based on numerical tests, this approximation is valid if the coupling is not too strong ($g/(E_1-E_0)\leq 0.2$), because in this case, nondegenerate basis states are not mixed significantly. With this approximation, the dark basis states are not coupled to any other basis state, while $\ket{G^{(1)}(n_0,N_{\rm mol}-n_0,0)}\ket{1}$ and $\ket{S^{(1)}(n_0-1,N_{\rm mol}-n_0,1)}\ket{0}$ form two-by-two blocks:
\begin{equation}
    \hat{\textbf{H}}_{\rm block}=
    \overset{\ket{G^{(1)}(n_0,N_{\rm mol}-n_0,0)}\ket{1} \quad \ket{S^{(1)}(n_0-1,N_{\rm mol}-n_0,1)}\ket{0}}
    {\begin{pmatrix}
        E(n_0-1,N_{\rm mol}-n_0,1)  &  g\sqrt{\frac{n_0}{N_{\rm mol}}}\mu    \\
        g\sqrt{\frac{n_0}{N_{\rm mol}}}\mu    & E(n_0-1,N_{\rm mol}-n_0,1)   \\
    \end{pmatrix}},\\
\end{equation}
where we indicated the basis state on the top of each column.
Therefore, the eigenstates of $\hat{H}$ (in this approximation) are the $\ket{D_k^{(i)}(n_0-1,N_{\rm mol}-n_0,1)}\ket{0}$ dark states with $E(n_0-1,N_{\rm mol}-n_0,1)$ energy, and the  
\begin{equation}
    \ket{P^{(1)}_{\pm}(n_0-1,N_{\rm mol}-n_0,1)}\equiv\frac{1}{\sqrt{2}}(\ket{G^{(1)}(n_0,N_{\rm mol}-n_0,0)}\ket{1}\pm \ket{S^{(1)}(n_0-1,N_{\rm mol}-n_0,1)}\ket{0})
\end{equation}
polaritonic states with $E(n_0-1,N_{\rm mol}-n_0,1)\pm g\sqrt{n_0/N_{\rm mol}}\mu$ energy.

\subsection{Spectrum calculation - Non-symmetric initial state}
Next, we obtain the spectrum for the case
where $n_0$ molecules are in $\ket{\psi_0}$ and $N_{\rm mol}-n_0$ molecules are in $\ket{\psi_1}$, and the initial states are the non-symmetric $\ket{G^{(i)}(n_0,N_{\rm mol}-n_0,0)}\ket{0}$ states. This scenario corresponds to, for example, a thermal molecular ensemble where the fraction of the molecules in $\ket{\psi_0}$ and $\ket{\psi_1}$ is determined by the Boltzmann-distribution (see main text).
First, we calculate the R-branch transitions, corresponding to the $\ket{\psi_0}\to \ket{\psi_2}$ excitation. The total intensity of the $\ket{0}\ket{G^{(i)}(n_0,N_{\rm mol}-n_0,0)}\to \ket{P^{(i)}_{\pm}(n_0-1,N_{\rm mol}-n_0,1)}$-type transitions to the polaritonic states is 
\begin{equation}
    I_{\rm R, polariton}(n_0,N_{\rm mol}-n_0,0)=\sum_{i=1}^{n_0}|\bra{0}\braket{G^{(i)}(n_0,N_{\rm mol}-n_0,0)|\hat{\mu}|P^{(i)}_{\pm}(n_0-1,N_{\rm mol}-n_0,1)}|^2=\mu^2 \frac{n_0}{2}{N_{\rm mol} \choose n_0},
\end{equation}
and the splitting of the peaks is $2g\sqrt{n_0/N_{\rm mol}}\mu$. On the other hand, R-transitions to the dark states, for example, $\ket{0}\ket{G^{(i)}(n_0,N_{\rm mol}-n_0,0)}\to \ket{D^{(i)}_k(n_0-1,N_{\rm mol}-n_0,1)}$, are forbidden:
\begin{equation}
    I_{\rm R, dark}(n_0,N_{\rm mol}-n_0,0)=\sum_{k=1}^{n_0-1}\sum_{i=1}^{n_0}|\bra{0}\braket{G^{(i)}(n_0,N_{\rm mol}-n_0,0)|\hat{\mu}|D^{(i)}_k(n_0-1,N_{\rm mol}-n_0,1)}|^2=0.
\end{equation}
Turning to the P-branch transitions, which corresponds to the $\ket{\psi_1}\to \ket{\psi_2}$ excitation, one can show that the total intensity of the  $\ket{0}\ket{G^{(i)}(n_0,N_{\rm mol}-n_0,0)}\to \ket{P^{(i)}_{\pm}(n_0,N_{\rm mol}-n_0-1,1)}$-type transitions to the polaritonic states is
\begin{equation}
    I_{\rm P, polariton}(n_0,N_{\rm mol}-n_0,0)=\sum_{i=1}^{n_0}|\bra{0}\braket{G^{(i)}(n_0,N_{\rm mol}-n_0,0)|\hat{\mu}|P^{(i)}_{\pm}(n_0,N_{\rm mol}-n_0-1,1)}|^2= \frac{\mu^2}{2}{N_{\rm mol} \choose n_0+1}
\end{equation}
and peak splitting is $2g\sqrt{(n_0+1)/N_{\rm mol}}\mu$. 
In contrast to the R-branch discussed above, the $\ket{0}\ket{G^{(i)}(n_0,N_{\rm mol}-n_0,0)}\to \ket{D^{(i)}_k(n_0,N_{\rm mol}-n_0-1,1)}$-type transitions to the dark states is allowed in the P-branch. The total intensity is\begin{equation}
    I_{\rm P, dark}(n_0,N_{\rm mol}-n_0,0)=\sum_{k=1}^{n_0}\sum_{i=1}^{n_0}|\bra{0}\braket{G^{(i)}(n_0,N_{\rm mol}-n_0,0)|\hat{\mu}|D^{(i)}_k(n_0,N_{\rm mol}-n_0-1,1)}|^2=\mu^2 n_0{N_{\rm mol} \choose n_0+1}.
\end{equation}
Therefore, the spectrum for finite $N_{\rm mol}$ becomes
\begin{equation}
\begin{aligned}
    I(\omega) = \sum_{+,-} \Big[ &
    n_0\frac{\mu^2}{2}{N_{\rm mol} \choose n_0}\delta\!\left(\omega_{02}\pm g\sqrt{\tfrac{n_0}{N_{\rm mol}}}\mu-\omega\right) \\
    & + \frac{\mu^2}{2}{N_{\rm mol} \choose n_0+1}\delta\!\left(\omega_{12}\pm g\sqrt{\tfrac{n_0+1}{N_{\rm mol}}}\mu-\omega\right) +
    n_0 \mu^2 {N_{\rm mol} \choose n_0+1}\delta\!\left(\omega_{12}-\omega\right) \Big],
\end{aligned}
\end{equation}
where the Dirac-delta represents the spectrum peaks and $\omega_{ij}=E_j-E_i$ are the cavity-free transition frequencies. The first term corresponds to the main resonant polariton peaks, the second term to the TP peaks, and the last to the transition to the dark states. There are two important points to be emphasized. First, the splitting of the main and the twin polariton peaks is not equal,  $2g\sqrt{n_0/N_{\rm mol}}\mu$ for the former and  $2g\sqrt{(n_0+1)/N_{\rm mol}}\mu$ for the latter. Of course, this difference diminishes in practice if both $n_0$ and  $N_{\rm mol}$ are large. Second, the intensity ratio of the P-branch transitions to the polaritonic states and to the dark states is $I_{\rm P, polariton}(n_0,N_{\rm mol}-n_0,0)/I_{\rm P, dark}(n_0,N_{\rm mol}-n_0,0)=1/(2n_0)$, which means that the peak of the dark state transitions completely suppresses the TP peaks in the thermodynamic limit where $n_0, N_{\rm mol} \to \infty$. 

Fig. 2(a) in the main text shows the calculated spectrum in the thermodynamic limit, $N_{\rm mol} \to \infty$, where $r_0=1/2$ fraction of the molecules are in $\ket{\psi_0}$, i.e., $n_0=r_0N_{\rm mol}=N_{\rm mol}/2$. Here, the spectrum shows only the primary polariton peaks for the R-branch (with $2g\sqrt{r_0}\mu$ splitting) and the dark-state transition for the P-branch, while the TP peaks disappear:
\begin{equation}
    I(\omega) \propto
    \sum_{+,-}\frac{\mu^2}{2}\delta\left(\omega_{02}\pm g\sqrt{r_0}\mu-\omega\right) + 
    \mu^2 \delta\left(\omega_{12}-\omega\right).
\end{equation}

\subsection{Spectrum calculation - Symmetric initial state}
In this section we obtain the transition intensities for initial states which are symmetric for the permutation of any molecule pair. In this example we take the initial state as $\ket{0}((\ket{\psi_0}+\ket{\psi_1})/\sqrt{2})^{\otimes N_{\rm mol}}$, which can be straightforwardly prepared in cold atom experiments and is relatively long lived.

Using a symmetric initial state allows the TP feature to survive in the thermodynamic limit, because transitions to the dark states will have zero intensity. Since both the initial state and the total dipole moment operator are symmetric for permutation of the molecules, the final state of the transition must be also symmetric, which is satisfied by the polariton states but not the dark states.

Let us assume that the initial state is a superposition state: 
\begin{equation}
    \ket{\Psi}=\sum_k c_k\ket{\Psi_k},    
\end{equation}
where $\ket{\Psi_k}$ is an eigenstate of $\hat{H}$ with $E_k$ energy. The spectrum, calculated form the dipole autocorrelation function, then becomes
\begin{equation}
    I(\omega)=\sum_{i,f} c^*_i\braket{\Psi_i|\hat{\mu}|\Psi_f}\left(\sum_m c_m\braket{\Psi_f|\hat{\mu}|\Psi_m}\right)\delta(\omega_{IF}-\omega),
\end{equation}
where the sums over $i$ and $f$ correspond to degenerate initial states and degenerate final states, with $\hat{H}\ket{\Psi_i}=E_I \ket{\Psi_i}$ and $\hat{H}\ket{\Psi_f}=E_F \ket{\Psi_f}$, respectively.

The symmetric initial state $\ket{0}((\ket{\psi_0}+\ket{\psi_1})/\sqrt{2})^{\otimes N_{\rm mol}}$ can be expressed by the ground state manifold states as
\begin{equation}\label{eq:superpos}
    \ket{0}(\ket{\psi_0}+\ket{\psi_1})/\sqrt{2})^{\otimes N_{\rm mol}} = \frac{1}{\sqrt{2^{N_{\rm mol}}}}\left(\sum_{i_1,...,i_{N_{\rm mol}}=0}^{1}\ket{i_1,...,i_{N_{\rm mol}}}\ket{0}\right)
\end{equation}

Then, using Eqs. (\ref{eq:mu_mtx_1}), (\ref{eq:mu_mtx_2}), and (\ref{eq:superpos}), one can show that the spectrum becomes
\begin{equation}
    I(\omega)=\frac{\mu^2}{2^{N_{\rm mol}}}\sum_{+,-}\sum_{n_0=0}^{N_{\rm mol}} {N_{\rm mol} \choose n_0} \left[n_0  \delta\left(\omega_{02} \pm g \mu \sqrt{\frac{n_0}{N_{\rm mol}}} - \omega\right)+(N_{\rm mol}-n_0)  \delta\left(\omega_{12} \pm g \mu \sqrt{\frac{n_0+1}{N_{\rm mol}}} - \omega\right)\right],
\end{equation}
where the first and second terms in the square brackets correspond to the primary polariton peaks (R-branch) and the TP (P-branch) peaks, respectively. In the thermodynamic limit, $N_{\rm mol} \to \infty$, one can show that the $n_0\approx N_{\rm mol}/2$ terms have the largest contribution, and the spectrum can be written as
\begin{equation}
    I(\omega)\propto \sum_{+,-}\left[\delta\left(\omega_{02} \pm g \sqrt{2} \mu  - \omega\right)+
    \delta\left(\omega_{12} \pm g \sqrt{2} \mu- \omega\right)\right].
\end{equation}

\vspace{0.1cm}
\bibliography{bibliogra}